\input{aipcheck}
\documentclass[
    ,draft            
  ]
  {aipproc}

\layoutstyle{6x9}

\begin{document}
\def\Msun{{M_\odot}}
\def\Zsun{{Z_\odot}}
\def\Lsun{{L_\odot}}

\title{The faintest galaxies}


\author{Stefania Salvadori}{
  address={Kapteyn Astronomical Institute, Landleven 12, 9747 AD Groningen, The Netherlands}
}

\author{Andrea Ferrara}{
  address={Scuola Normale Superiore, Piazza dei Cavalieri 7, 56126 Pisa, Italy}
}

\begin{abstract}
We investigate the nature of Ultra Faint dwarf spheroidal galaxies (UF dSphs) 
in a general cosmological context, simultaneously accounting for various 
``classical'' dSphs and Milky Way (MW) properties, including their Metallicity 
Distribution Function (MDF). The model successfully reproduces both the observed 
[Fe/H]-Luminosity relation and the mean MDF of UFs. According to our results UFs 
are the living fossils of H$_2$-cooling minihaloes formed at $z>8.5$, i.e. before
the end of reionization. They are the oldest and the most dark matter-dominated 
($M/L > 100$) dSphs in the MW system, with a total mass of $M = 10^{7-8}\Msun$. 
The model allows to interpret the different shape of UFs and classical dSphs MDF, 
along with the frequency of extremely metal-poor stars in these objects. 
We discuss the ``missing satellites problem'' by comparing the UF star formation 
efficiencies with those derived for minihaloes in the Via Lactea simulation. 
\end{abstract}
\maketitle
Ultra faint dwarf spheroidal galaxies (UF dSphs) represent the {\it least} luminous
($L< 10^5L_{\odot}$), the {\it least} metal-rich ($\langle$[Fe/H]$\rangle<-2$) 
and probably the {\it least} massive ($M<10^8 M_{\odot}$) stellar systems ever 
known. Such extreme features make these galaxies the best suitable living fossils 
for the investigation of the early cosmic star formation.  
In addition, recent surveys of very metal-poor stars in dSphs pointed out
that [Fe/H]$<-3$ stars represent the $25\%$ of the total stellar mass in UFs
\cite{Kirby}, while they are greatly rare \cite{Else} in ``classical'' dSphs.
When do UFs form? Could these galaxies represent the first star-forming objects
in the MW system?

In this contribution we investigate the nature of UFs from a general 
cosmological prospective. To this aim we use the semi-analytical code GAMETE 
(GAlaxyMErgerTree\&Evolution), which traces the hierarchical build-up of the 
Milky Way (MW) system, simultaneously accounting for various classical dSphs 
and MW properties, including their Metallicity Distribution Functions (MDFs). 
\begin{figure}
  \includegraphics[height=.25\textheight]{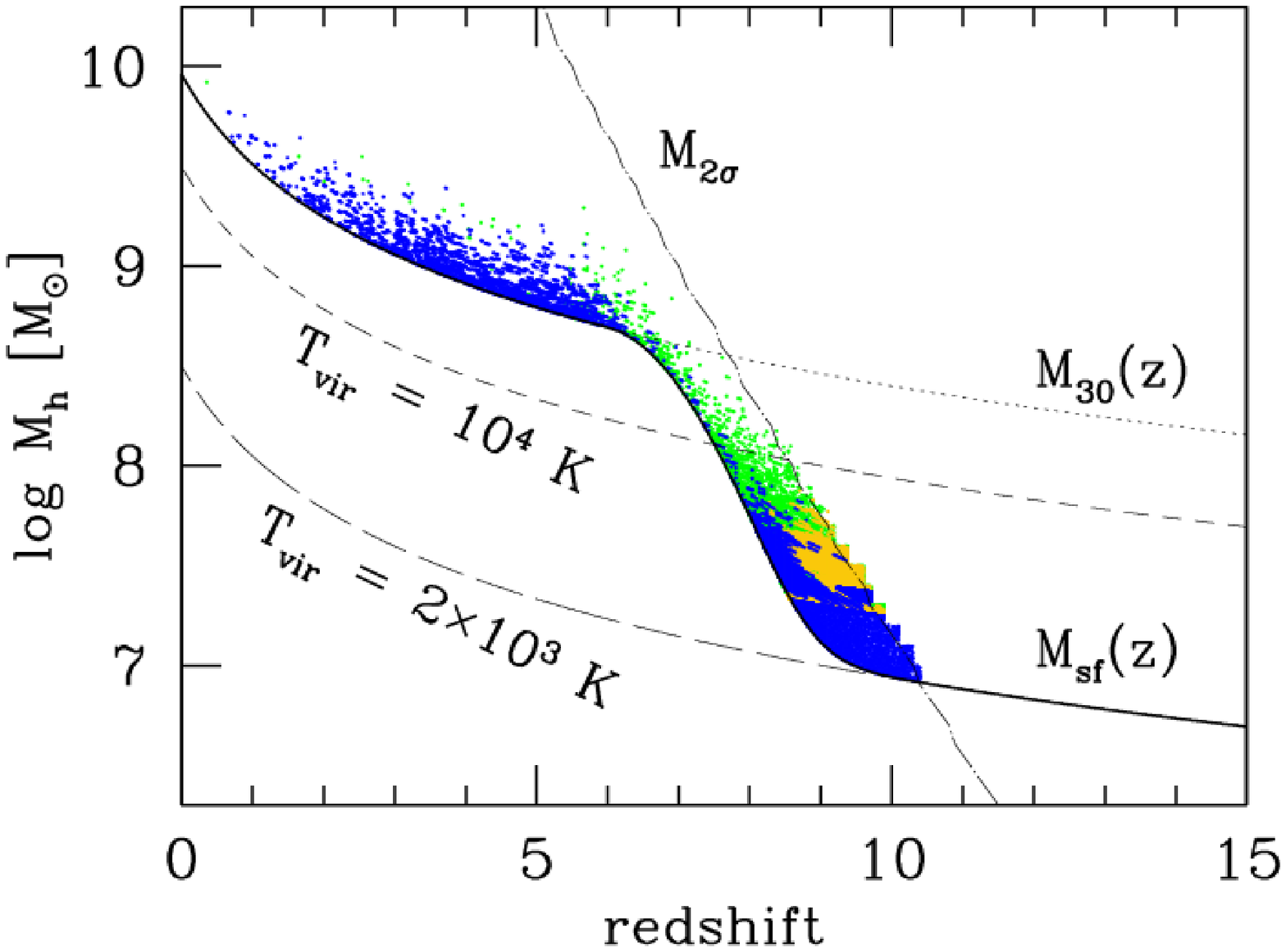}
  \includegraphics[height=.25\textheight]{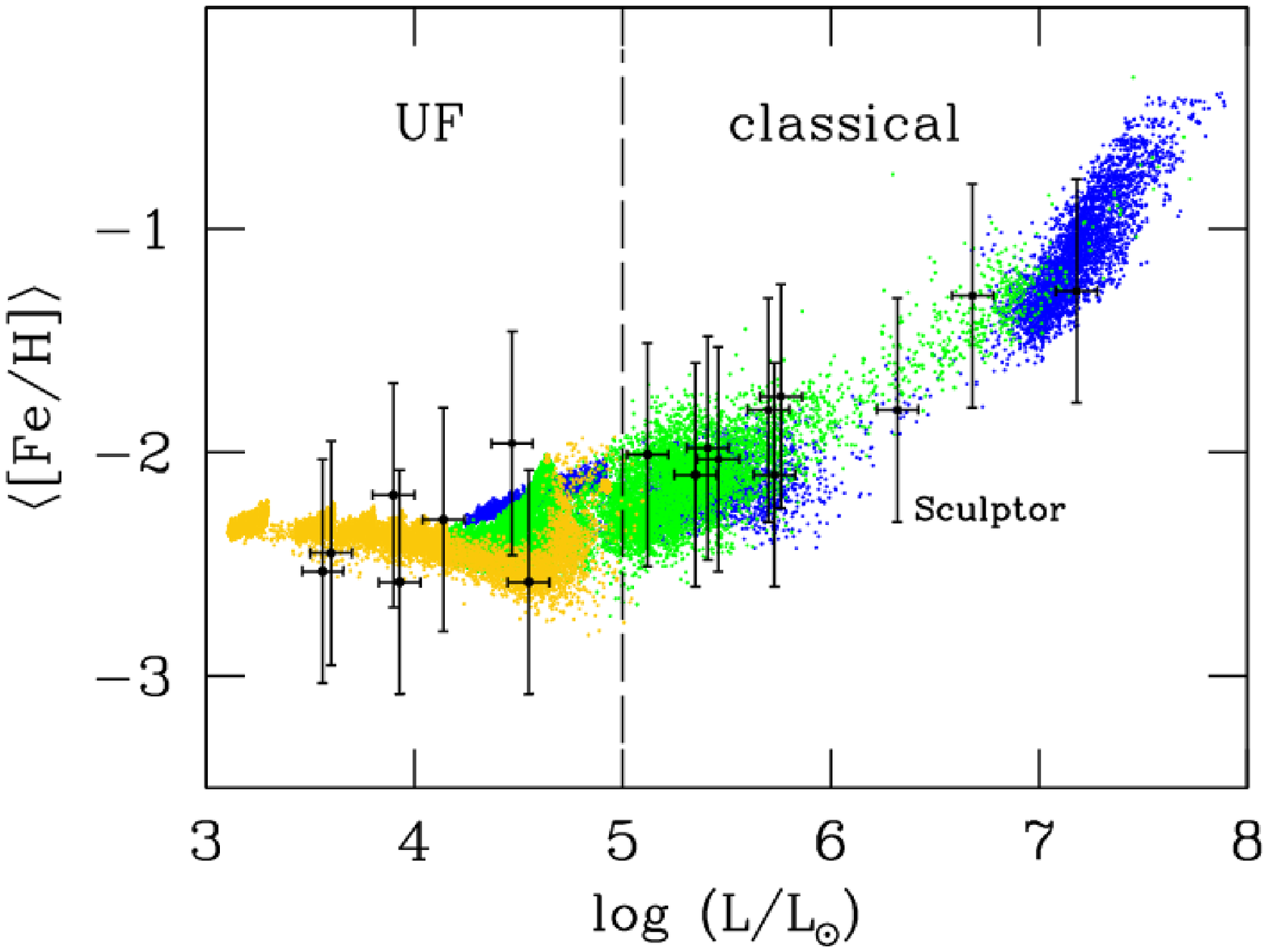}
  \caption{Selected dSph candidates (points) in 10 possible merger histories. 
    Different colors show the baryonic fraction $f_b$ at the formation epoch 
    with respect to the cosmic value $f_c=0.156$: $f_b/f_c > 0.5$ (blue), 
    $0.1 < f_b/f_c < 0.5$ (green), $f_b/f_c < 0.1$ (yellow). We show: 
    {\it Panel a:} the dSph hosting halo mass and circular velocity as a 
    function of the formation redshift $z$.
    The lines in the panel show the evolution of $M_{sf}(z)$ (solid), 
    the halo mass corresponding to 2$\sigma$ peaks (dotted-long dashed),
    $T_{vir}=10^4$K (short dashed line) and $T_{vir}=2\times 10^3$K (long dashed
    line). {\it Panel b:} the total luminosity of dSphs as a function of their 
    iron-abundance (points). Points with error bars are observational data 
    (\cite{Kirby}).}
  \label{fig:Fe_L}
\end{figure}
\subsection{Star formation and feedback processes}   
The basic features of the model can be summarized in few points (for a complete 
description see \cite{SSF07,SFS08,SF09}). After having reconstruct a statistical 
significant sample of MW hierarchical merger histories we follow the evolution 
of gas and stars along each hierarchical tree by assuming that:
(i) stars can only form in objects above a minimum halo mass, $M_{sf}(z)$, whose 
evolution (Fig.~1, left panel) accounts for the suppression of star formation (SF) 
in progressively more massive objects due to radiative feedback effects \cite{SF09};
(ii) the reionization of the MW environment is complete at $z_{rei}=6$;
(iii) in each galaxy the SF rate is proportional to the mass of cold gas, whose
gradual accretion is regulated by a numerically calibrated infall rate \cite{Keres};
(iv) due to ineffective cooling by H$_2$ molecules the SF efficiency in ``minihaloes'' 
($T_{vir}<10^4$~K) is reduced $\propto[1+(T_{vir}/2\times 10^4{\rm K})^{-3}]^{-1}$
with respect to H-cooling haloes.\\
The chemical enrichment of the gas is followed both in proto-Galactic haloes and in 
the MW environment by taking into account the mass-dependent stellar evolutionary 
timescales and the effects of mechanical feedback due to supernova (SN) energy 
deposition \cite{SFS08}. The free parameters of the model (SF and SN wind 
efficiencies) are calibrated to {\it simultaneously reproduce \cite{SSF07} the 
global properties of the MW} (stellar/gas mass and metallicity) {\it and the MDF 
of Galactic halo stars} \cite{Schoerck,S10}. 
\subsection{The satellite candidates}
The dSph candidates are selected among the star forming haloes ($M > M_{sf}(z)$) which are 
likely to become satellites i.e. those corresponding to density fluctuations $<2\sigma$ 
\cite{Diemand}. Their total dark matter (DM) mass, formation redshift and initial baryonic 
fraction $(f_b$) with respect to the cosmic value $f_c=\Omega_b/\Omega_m=0.156$ are shown 
in Fig.~1 (left panel). Gas-rich systems, $f_b/f_c > 0.5 $, are newly virialized objects 
accreting gas from the MW environment; intermediate (gas-poor) systems, 
$0.1 < f_b/f_c < 0.5$ ($f_b/f_c < 0.1$), originate from the mixed merging of star-forming 
(newly virializing) and starless progenitors; their smaller baryonic content is the result 
of shock-heating of the infalling gas during major merging events \cite{Cox}. 
According to our results dSph galaxies are hosted by DM haloes with masses 
$M\approx10^{7-9}\Msun$, virialized out of the MW environment between $z\approx 10.5$ 
and $z\approx 3$.
\begin{figure}
  \includegraphics[height=.30\textheight]{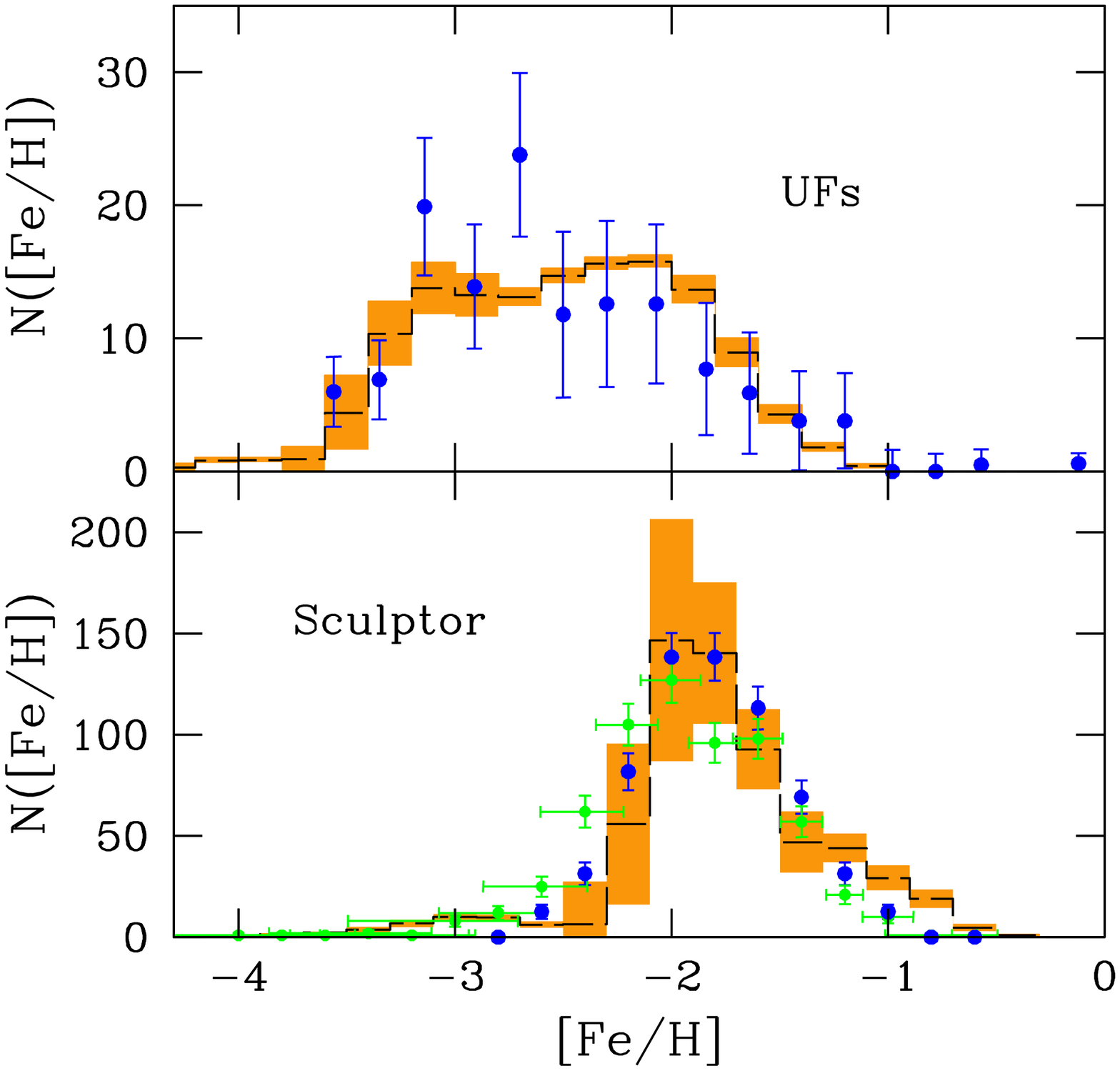}
  \includegraphics[height=.28\textheight]{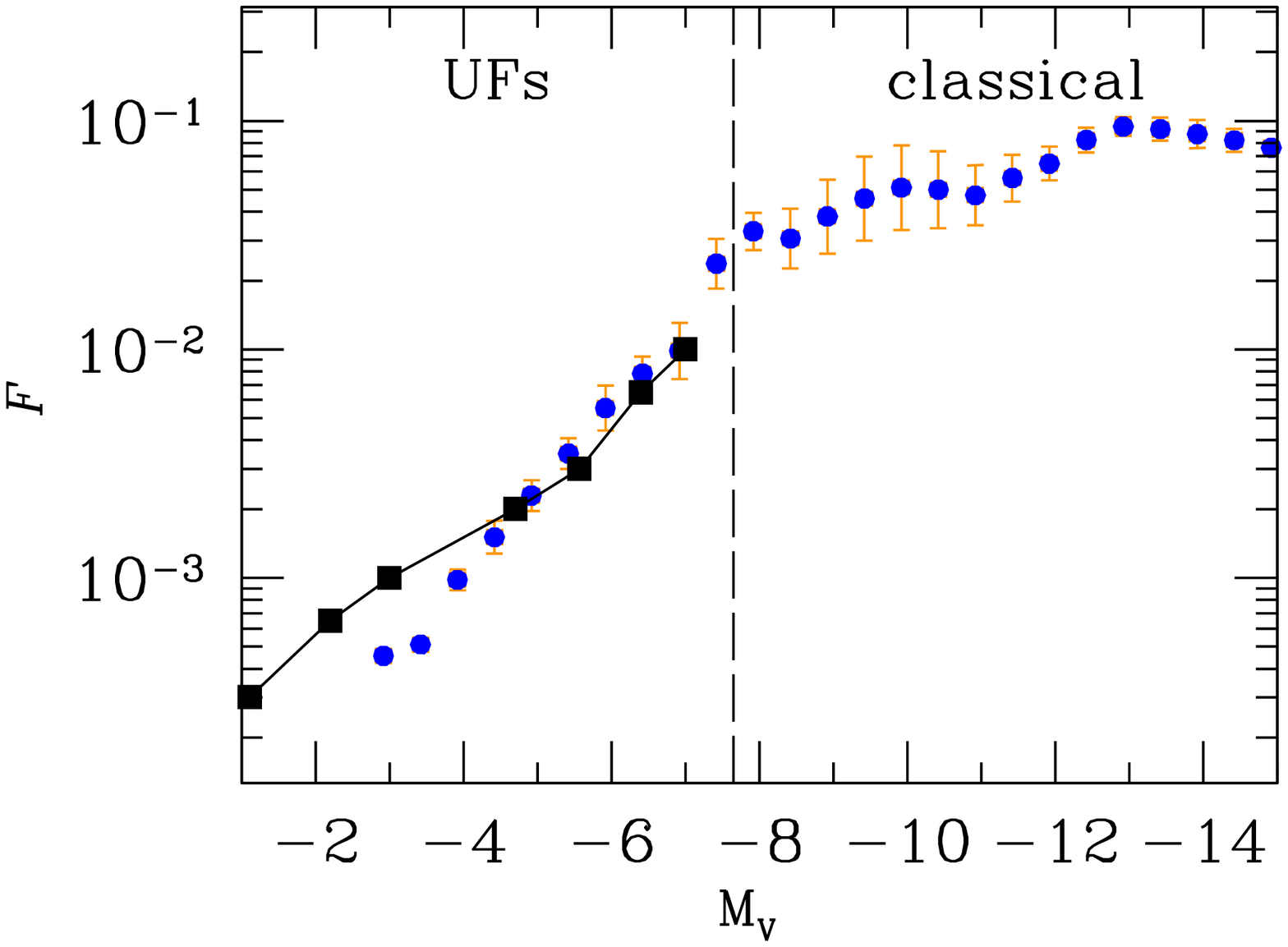}
    \caption{{\it Left panels:} observed (points) and simulated (histogram)
    MDF of UFs ({\it top}) and Sculptor dSph ({\it bottom}). Histograms are
    the averaged MDF over all UFs ($L<10^5L_{\odot}$ {\it top}) and Sculptor 
    ($10^6 L_{\odot}<L<10^{6.5} L_{\odot}$ {\it bottom}) candidates in 10 
    merger histories. The shaded area is the $\pm 1\sigma$ scatter among 
    different realizations. The data points are by \cite{Kirby} ({\it top}, 
    Poissonian errors) and by the DART team ({\it bottom}) using the old 
    (blue points, Poissonian errors \cite{Helmi}) and the new (green points, 
    observational errors \cite{Else}) CaT line calibration.
    {\it Right panel:} fraction ${\cal{F}} = M_*/f_c M$ of the potentially
    available cosmic baryons turned into stars as a function of $M_V$. 
    The points are the average over all dSph candidates in 10 realizations. 
    Error bars show the $\pm 1\sigma$ dispersion among different dSphs. 
    The squared points are the results by \cite{Madau}}
\label{fig:MDFs}
\end{figure}
\subsubsection{The Fe-luminosity relation}
The observed Fe-Luminosity relation of dSph galaxies is well reproduced by the
model (Fig.~1, right panel). We found that minihaloes predominantly populate the 
faint end of the relation, $L<10^6\Lsun$ while above that luminosity H-cooling 
haloes dominates. We therefore conclude that {\it all UFs are left-overs of 
H$2$-cooling minihaloes}, in good agreement with the findings by \cite{Bovill}. 
As minihaloes virialized when $z > 8.5$ and have total masses $M\approx 10^{7-8}\Msun$ 
(left panel), we infer that {\it UFs are the oldest and the more DM dominated ($M/L>100$) 
dSphs in the MW system}. In the faintest UFs, $L<10^4\Lsun$, which are gas poor 
systems, the mass-to-light ratio reach extreme values such as $M/L \approx 10^4$.
\subsubsection{Metallicity Distribution Functions}
The observed and simulated MDFs of UFs (\cite{Kirby}) and Sculptor dSph 
(\cite{Helmi,Else}) are compared in Fig.~2 (left panels). The UFs MDF is 
broader with respect to the Sculptor MDF because of their more prolonged 
SF history \cite{SF09}, while it is shifted towards lower [Fe/H] values 
as a result of the lower metallicity of the MW environment at the time of 
their formation: Sculptor-like dSphs are associated with gas-rich, H-cooling 
haloes, that virialize at $z\approx 7.5$, when the {\it pre-enrichment} of 
the MW environment was [Fe/H]$\approx -3$ (\cite{SFS08}); UFs instead form 
at higher redshifts ($z>8.5$) when the metallicity was lower.
Note the small [Fe/H]$< -3$ tail in Sculptor predicts by the model and now 
confirmed by the new results of the DART survey \cite{Else} (see the Figure) 
and by high resolution spectroscopic studies \cite{Frebela}. In our picture 
these stars are found to be stellar relics of the rare SF episodes occurred 
in some progenitor minihaloes, when $z > 7.5$. This implies that [Fe/H]$<-3$ 
stars in classical dSphs are expected to have the same abundance pattern 
of that in UFs \cite{SF09}. Recent high-resolution analysis of extremely 
iron-poor stars in Sculptor \cite{Frebelb} and UFs \cite{Frebela,Koch} 
have confirmed this prediction. 
\subsubsection{Are there missing satellites?}
The fraction of the potentially available cosmic baryons turned into stars by 
$z=0$ (${\cal{F}} = M_*/f_c M$) is shown in Fig.~3, as a function of the dSph 
magnitude, $M_V$. Model results are compared with the findings of \cite{Madau}, 
which determine ${\cal F}$ for minihaloes by matching the luminosity function 
of the MW satellites in the SDSS. The agreement between the two studies is very  
good. Hence, although our method prevents us from making specific predictions on 
the actual number of satellites, we conclude that {\it there is no missing 
satellites problem in terms of SF efficiency}.
\subsection{Conclusions}
Ultra Faint dSphs are the today-living counterpart of high redshift, H$2$-cooling 
minihaloes. They are the oldest ($z>8.5$) and more DM dominated ($M/L >100$) 
galaxies in the MW system. They are very ineffectively SF objects, turning into 
stars $<3\%$ of the potentially available cosmic baryons. The higher fraction 
of [Fe/H]$<-3$ stars in UFs, with respect to classical dSphs, reflects both 
their lower SF rate, caused by ineffective H$_2$ cooling, both the lower metal 
(pre-)enrichment of the MW-environment at their (further) formation epoch.
At the moment, these faint galaxies represent the best available living fossil 
to investigate the early cosmic star formation.
\bibliographystyle{aipproc}
\bibliography{sample}

\end{document}